\documentclass[amsmath,twocolumn,superscriptaddress,prl,aps]{revtex4}
\usepackage{amsthm,amsfonts,graphicx,verbatim}
\usepackage{bm}% bold math
\newcommand{\be}{\begin{equation}}
\newcommand{\ee}{\end{equation}}
\newcommand{\bea}{\begin{eqnarray}}
\newcommand{\eea}{\end{eqnarray}}

\newcommand{\Tr}{{\rm Tr}\,}
\renewcommand{\phi}{\varphi}
\renewcommand{\epsilon}{\varepsilon}
\renewcommand{\vec}[1]{{\bf #1}}

\renewcommand{\cite}[1]{[\onlinecite{#1}]}

\newcommand{\sign}{\mathrm{sign}}

\begin{document}

\title{
An itinerant half-metal spin-density-wave state on the hexagonal lattice}
 \author{Rahul Nandkishore}
 \affiliation{Department of Physics, Massachusetts Institute of Technology, Cambridge MA 02139, USA}
\author{Gia-Wei Chern}
\affiliation{Department of Physics, University of Wisconsin-Madison, Madison, WI 53706, USA}
\author{Andrey V. Chubukov}
\affiliation{Department of Physics, University of Wisconsin-Madison, Madison, WI 53706, USA}

\begin{abstract}
We consider electrons on a honeycomb or triangular lattice doped to the saddle point of the bandstructure. We assume system parameters are such that spin density wave (SDW) order emerges
 below a temperature $T_N$
 and investigate the nature of the SDW phase. We
 argue that at $T \leq T_N$
  the system develops a  uniaxial SDW phase
   whose ordering pattern
    breaks $O(3) \times Z_4$ symmetry and
    corresponds to  an eight site unit cell with non-uniform spin moments on different sites.
  This state is a half-metal --
  it preserves full original Fermi surface, but has
  gapless charged excitations
  in one spin branch only.
   It allows for electrical control of spin currents
     and  is desirable for nano-science.

\end{abstract}

\maketitle
{\it Introduction:}~~~~The electronic properties of single layer graphene have been the subject of considerable experimental and theoretical interest \cite{CastroNeto}.
 Near half-filling, a description in terms of non-interacting Dirac electrons captures the essential
  physics,
   since interactions effects are suppressed by the low density of states (DOS). A sharply different behavior
 arises when graphene is strongly doped to
   $3/8$ or $5/8$ filling \cite{McChesney}.
  At this filling, a divergent density of states and nested Fermi surface (FS) conspire to produce weak coupling instabilities to an extensive buffet of
   ordered states, including
  spin density waves (SDW) \cite{TaoLi, MoraisSmith,FaWang}, Pomeranchuk metals \cite{Vozmediano}, and $d$ wave superconductors (SC)
   \cite{Gonzalez,Nandkishore,Thomale}.
 A similar situation arises on a triangular lattice at $3/4$ filling~\cite{Martin+Batista,TaoLioriginal}.

It has recently been established using renormalization group (RG) methods \cite{Nandkishore} that the two most relevant instabilities at
 weak coupling are towards SDW and
 a $d$-wave SC.
 The SDW vertex  is the largest at intermediate RG scales, but superconducting vertex eventually
 overshoots it both at perfect nesting and away from perfect nesting,
   making  $d$-wave superconductivity
  the leading  weak coupling instability.
  The SC state has a
 $d+id$ gap structure and breaks time-reversal symmetry \cite{Nandkishore}.

   In this paper we assume that superconductivity is destroyed by an applied  magnetic field, or
   alternatively that system parameters are such that corrections to the RG flow become relevant before SC vertex overshoots SDW vertex.  In both cases, the SDW instability becomes dominant, and an SDW order emerges.  Previous work argued that the SDW state is non-coplanar and has a non-zero spin chirality \cite{TaoLi,FaWang}.  Such a state gaps out the entire Fermi surface (FS), i.e., is an insulator. The chiral SDW state has also been found in the strong coupling analysis for classical spins of fixed length \cite{Martin+Batista}.

\begin{figure}
a) \includegraphics[width = 0.4\columnwidth]{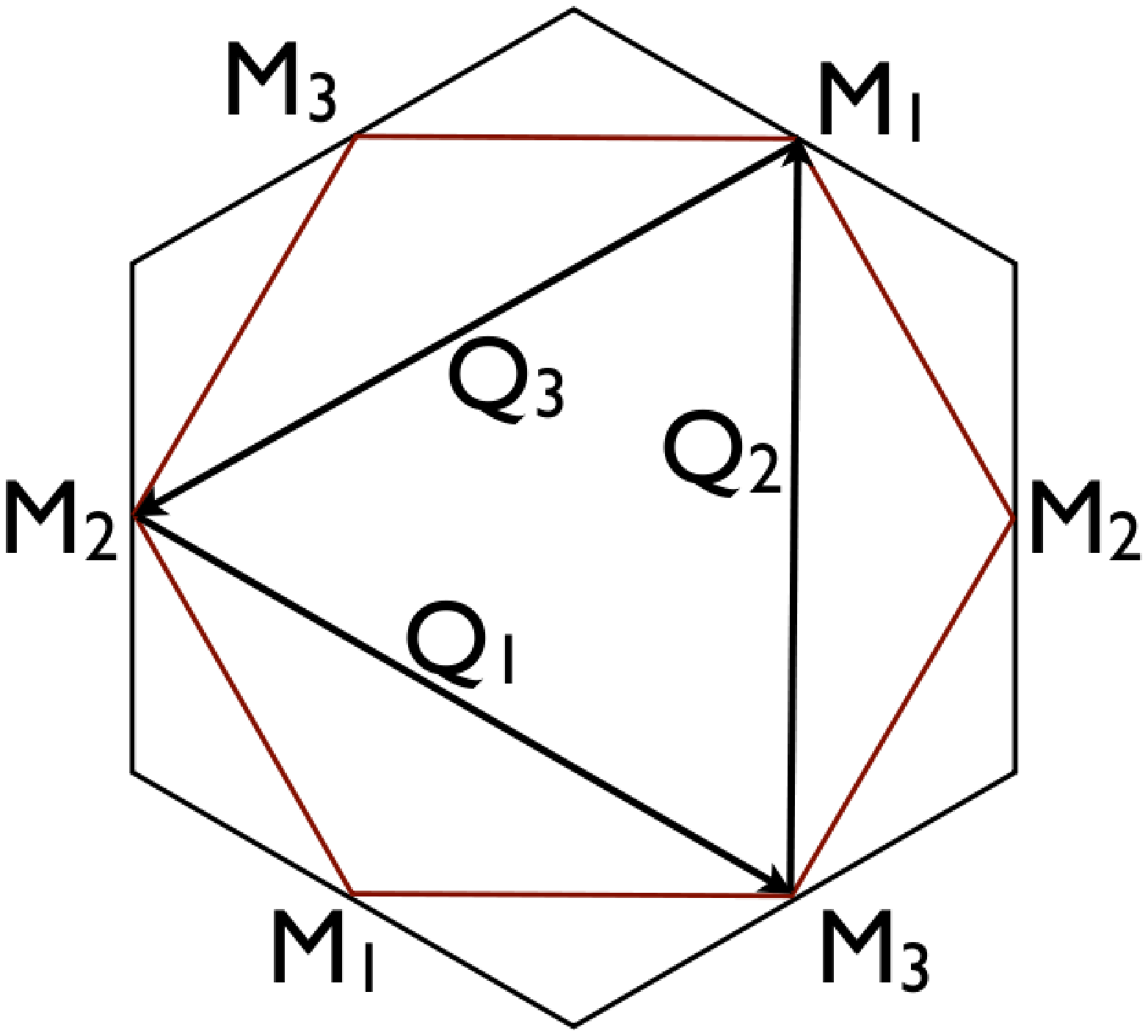}
b) \includegraphics[width = 0.47\columnwidth]{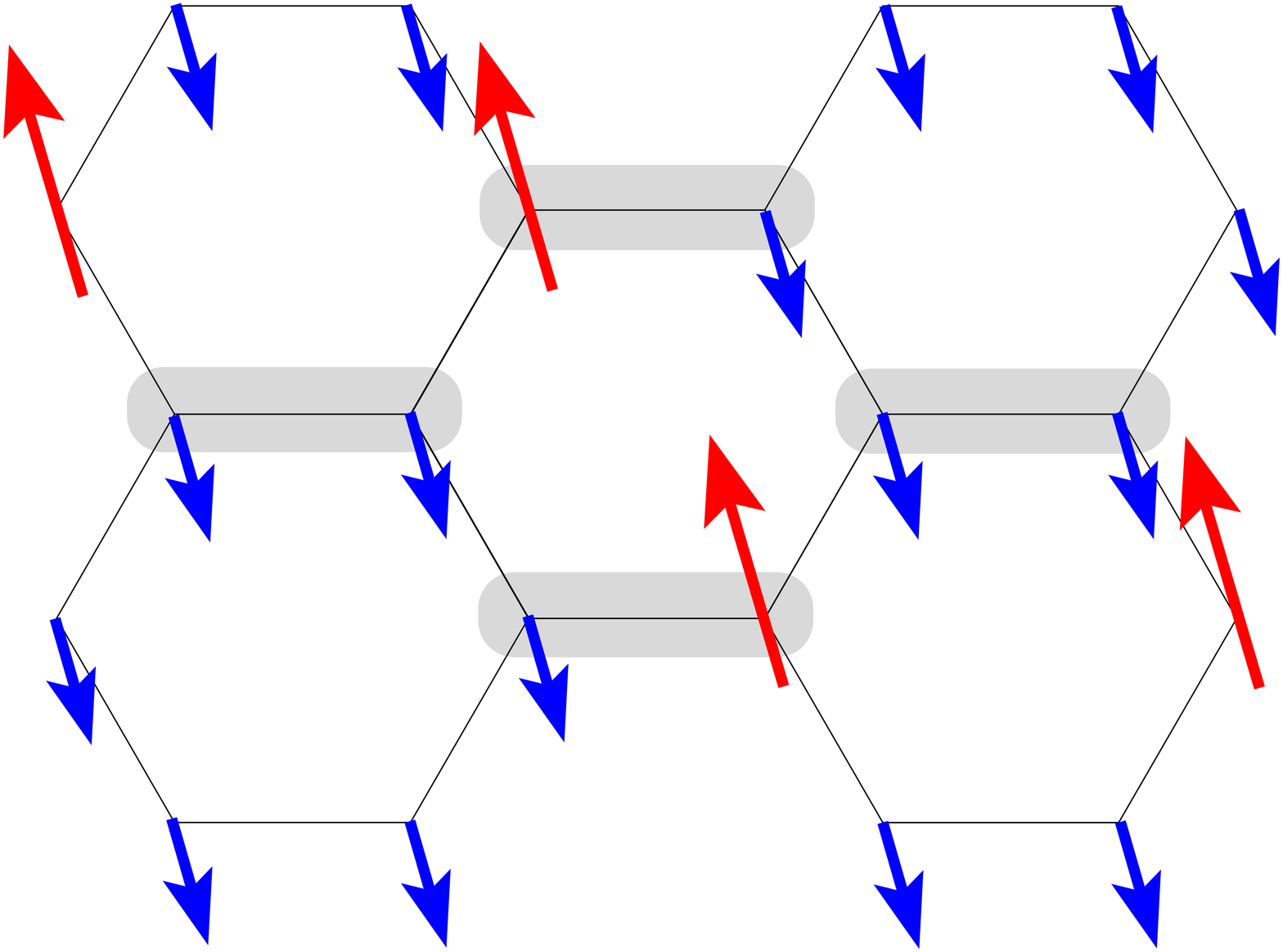}
\caption{(Color online) a) The Fermi surface at the doping level of interest is a hexagon inscribed within a hexagonal Brillouin zone
 (BZ),
 for both honeycomb and triangular lattices. The FS has three inequivalent corners, which are saddle points of the dispersion, marked by a vanishing Fermi velocity and a divergent density of states. The three inequivalent saddle points $M_i$ are connected by three inequivalent nesting vectors $\vec{Q}_i$, each of which is equal to half a reciprocal lattice vector, such that $\vec{Q}_i = - \vec{Q}_i$. b)
 Spin structure for the uniaxial
  SDW state.
  The SDW order quadruples the unit cell to a unit cell with eight sites
   (shaded).
    The enlarged unit cell has a large spin moment $3 \Delta$ on two sites and a small spin moment $-\Delta$ on the other six. The total spin on each unit cell
     is
     zero.
  \label{fig: lattice}}
\end{figure}

We argue that the situation is more complex than originally thought,
   and
   the chiral SDW state is present only at the lowest temperatures. Over a wide intermediate range of temperatures,
    a different SDW state emerges in which SDW
    order
    develops simultaneously at three inequivalent wavevectors ${\mathbf Q}_i$, but the three vector order parameters are all aligned along the same axis.
 This state has a eight site unit cell with non-uniform spin moments and zero net magnetization (Fig.~\ref{fig: lattice}b). Such a state cannot be accessed starting from a spin Hamiltonian for local moments
   with a fixed length,
    and can only be accessed starting from a model of itinerant fermions. We show that in this state, unlike in any other known SDW state, the chemical potential shifts proportionally to the
     SDW order parameter  preserving the original Fermi surface for  one spin branch and gapping out the other spin branch.  The uniaxial SDW state
    is therefore a `half-metal' that allows for electrical control of spin currents.
     Such a state is highly desirable for nano-science applications.

{\it The model:}~~~~For definiteness we focus on doped graphene at
 $3/8$ filling.
 Our point of departure is the tight binding model \cite{Wallace},
  with the nearest-neighbor dispersion
\begin{equation}
 \label{eq: dispersion}
\epsilon_{\vec{k}} = -t_1 \sqrt{1 + 4 \cos{\frac{k_y \sqrt{3}}{2}} \cos{\frac{3 k_x}{2}} + 4 \cos^2{\frac{k_y \sqrt{3}}{2}}} - \mu
\end{equation}
where $\mu = -t_1$ at $3/8$ filling.
 The FS then
 forms a perfect hexagon inscribed within a hexagonal
  BZ (Fig.~\ref{fig: lattice}a).
   The perfect nesting of the FS in doped graphene is quite robust -- it is
   broken only by third and higher neighbor hoppings, which are generally quite small.
    The Fermi velocity vanishes near the hexagon corners ${\bf M}_1 = (2\pi/3,0),~{\bf M}_2 = (\pi/3,\pi/\sqrt{3}),~{\bf M}_3 = (\pi/3,-\pi/\sqrt{3})$,  which are saddle points of the dispersion:
 \begin{equation}
\epsilon_{{\bf M}_1+\mathbf k} \approx \frac{3t_1}{4}(k^2_y - 3k_x^2),~~\epsilon_{\mathbf M_{2,3}+\mathbf k} \approx -\frac{3t_1}{4} 2k_y(k_y \mp \sqrt{3}k_x),
\label{n_1}
\end{equation}
 where
  each time
 $\mathbf k$ denotes the deviation from a saddle point.
  Saddle points give rise to a logarithmic singularity in the DOS and control the SDW instability at weak coupling.
  There are three in-equivalent nesting vectors
  connecting in-equivalent pairs of saddle points
(see Fig.~\ref{fig: lattice}a):
 \begin{equation}
\mathbf Q_1 = (0,2\pi/\sqrt{3}) , ~~\mathbf Q_{2,3} = (\pm \pi/3, - \pi/\sqrt{3}).
\end{equation}
 Each $\vec{Q}_i$ is equivalent to $-\vec{Q}_i$ modulo a reciprocal lattice vector.

  There are two
   electron-electron
   interactions that
    contribute to the SDW channel. One is a forward scattering interaction $|\vec{k}, \vec{k}+\vec{Q}_i\rangle \rightarrow |\vec{k}, \vec{k}+\vec{Q}_i\rangle$, while the other is an umklapp interaction, $|\vec{k}, \vec{k'}\rangle \rightarrow |\vec{k}+ \vec{Q}_i, \vec{k'}+\vec{Q}_i\rangle$. We label these interactions $g_2$ and $g_3$ respectively, for consistency with the notation introduced in \cite{Nandkishore}.
The  partition function
 for $g_2-g_3$ model
can then be written as $Z = \int D[\psi^{\dag},\psi] \exp(- S[\psi^{\dag}, \psi])$, where $S = \int_0^{1/T} \mathcal{L}(\vec{k},\tau)$ and
\begin{eqnarray}
\mathcal{L} &=& \sum_{\alpha} \psi^{\dag}_{a,\alpha} (\partial_{\tau} - \epsilon_{\vec{k}} + \mu) \psi_{a,\alpha} \nonumber\\ &-& \sum_{\alpha \neq \beta} g_3 \psi^{\dag}_{a,\alpha}\psi^{\dag}_{a,\beta}  \psi_{b,\beta} \psi_{b,\alpha} - g_2 \psi^{\dag}_{a,\alpha}\psi^{\dag}_{b,\beta}
 \psi_{b,\beta} \psi_{a, \alpha}, \quad  \label{eq: sdw action}
\end{eqnarray}
where the action is written in terms of electron operators, $a, b$ are patch labels, and $\alpha$ and $\beta$ are spin components.

   Each nesting vector ${ \vec Q}_{i}$ has associated with it an SDW
    order parameter ${\vec \Delta}_{i} = {\vec \Delta}_{a,b} = \frac{g_2+g_3}{3}\sum_k \langle \phi^\dagger_{a,\alpha} {\bm \sigma}_{\alpha \beta} \phi_{b,\beta}\rangle$.
    The condition for the emergence of each ${\vec \Delta}_i$ is the same:
      $((g_2 + g_3)/t_1) \log^2 t_1/T_N = O(1)$ \cite{Nandkishore},
     leaving a large number of SDW states as potential candidates. We study the selection of the SDW order
       within
       Ginzburg-Landau theory and by comparing different SDW solutions in the mean-field approximation for
        Eq.~(\ref{eq: sdw action}) at arbitrary $T < T_N$.

 {\it Ginzburg-Landau theory:}~
 To construct  the Ginzburg-Landau theory, we decouple the quartic interaction terms by restricting the interaction to the spin channel and performing a Hubbard Stratonovich  transformation to introduce the order parameters ${\vec \Delta}_{i}$. We integrate out the fermions in the Matsubara frequency representation and obtain an action in terms of the order parameter fields ${\vec \Delta}_i$, which takes the form
 \begin{eqnarray}
\mathcal{L} &=& T \sum_{n=-\infty}^{\infty} \int \frac{d^2 k }{(2\pi)^2} \bigg[\frac{2}{g_2+g_3} \sum_{i}    ({\vec \Delta}_i)^2 \nonumber\\
&+&  \Tr \ln \bigg(i\omega_n - \epsilon_\vec{k} - \sum_{i}
{\vec \Delta}_{i}\cdot {\bm \sigma}\bigg) \bigg]. \label{eq: grand potential}
\end{eqnarray}

\begin{figure}
\includegraphics[width = 0.8 \columnwidth]{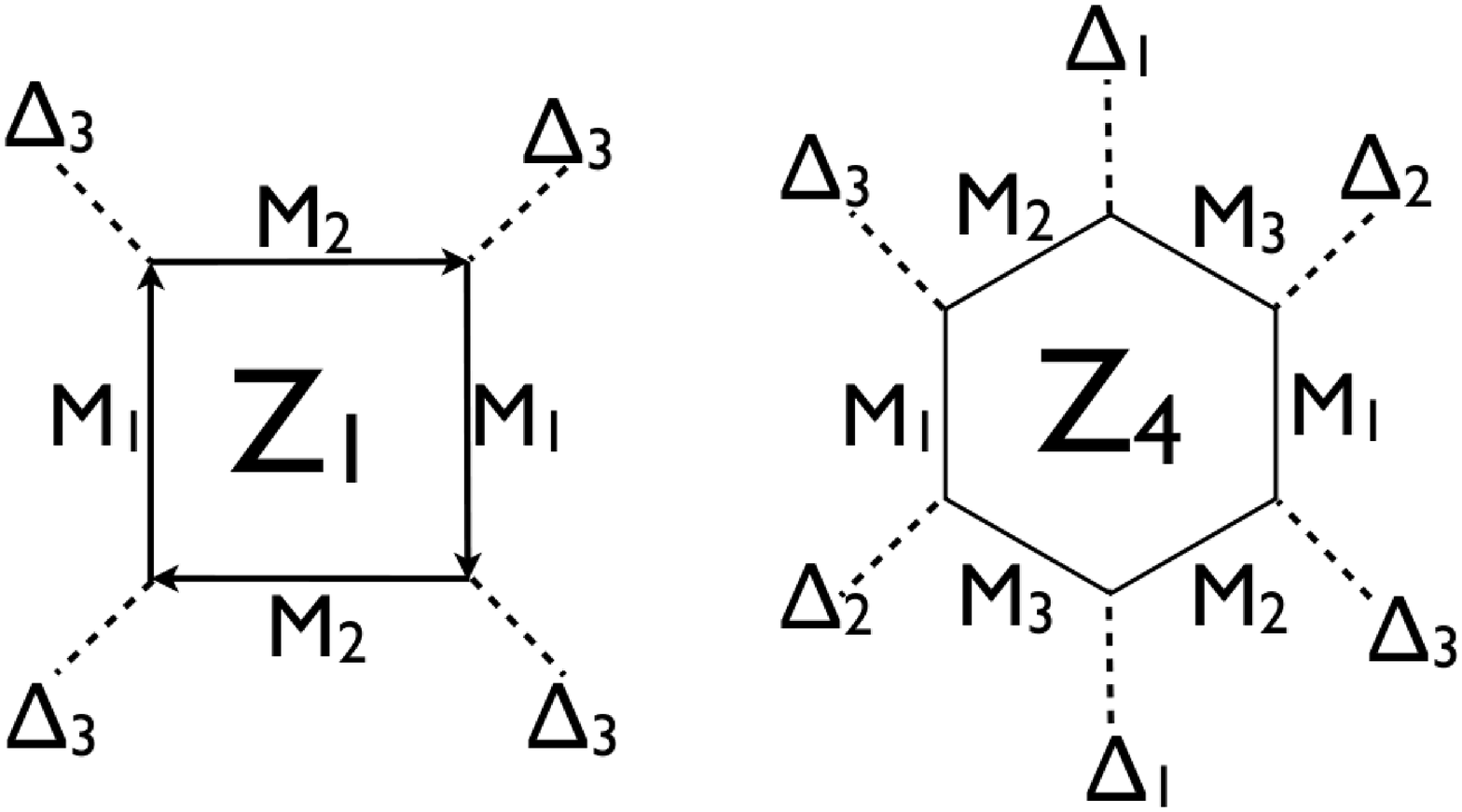}
\caption{(Color online) The terms quartic in $\Delta$ are produced by processes represented diagrammatically by square diagrams.
 The diagrams for $Z_2$ and $Z_3$ correspond to patterns $\Delta_3, \Delta_3, \Delta_1, \Delta_1$ and $\Delta_3, \Delta_1, \Delta_3, \Delta_1$, respectively.
  The sixth order chirality sensitive term is produced by `hexagonal diagrams.' Sample square and hexagonal diagrams are shown above. The integrals are dominated by momenta that bring all the fermion propagators to the vicinity of one of the saddle points of the dispersion. \label{fig: diagrams}}
\end{figure}

For $T \approx T_N$, we can expand (\ref{eq: grand potential}) in small ${\vec \Delta}_i/T_N$. It is useful to define the expansion coefficients
 \begin{equation}
  Z_i = T \sum_{\omega_n} \int \frac{d^2k}{(2\pi)^2} \xi_i
  \end{equation}
  where the integrands $\xi_i$ are expressed in terms of fermionic Green functions $ G = (i\omega_n - \epsilon_{\vec{k}}-\mu)^{-1}$,
   $G_i =(i\omega_n - \epsilon_{\vec{k+Q}_i}-\mu)^{-1}$, and
    $G_{i+j} = (i\omega_n - \epsilon_{\vec{k+Q}_i+\vec{Q}_j}-\mu)^{-1}$ as
   \begin{eqnarray}
\xi_1& =& G^2 G^2_3, \quad \qquad \qquad \qquad \xi_2 = G^2 G_3 G_1,~~\nonumber\\
\xi_3 &=&  G G_3 G_1 G_{1+3},\qquad \qquad
\xi_4 = G^2 G^2_3 G^2_1.
\label{eq: Z4}
\end{eqnarray}
  Diagrammatically, $Z_1$ --$Z_3$ are given by `square' diagrams with four fermionic propagators and ${\bm \sigma}_{\alpha \beta}$ in the vertices,
   and $Z_4$ is given by a `hegagonal' diagram with six fermionic propagators,
   (see Fig.~\ref{fig: diagrams}).
The free energy evaluated at $T\approx T_N$ can be expressed in terms of these coefficients as

\begin{widetext}
\begin{eqnarray}
\mathcal{L}
& \propto& \alpha (T-T_N) \sum_{i} {\vec \Delta}_{i}^2 +  Z_1 ({\vec \Delta}_{1}^2 + {\vec \Delta}_{2}^2 + {\vec \Delta}_{3}^2 )^2
+ 2 (Z_2-Z_1-Z_3) ({\vec \Delta}_{1}^2 {\vec \Delta}_{2}^2 + {\vec \Delta}_{2}^2 {\vec \Delta}_{3}^2 + {\vec \Delta}_{3}^2 {\vec \Delta}_{1}^2) \nonumber\\
&+& 4 Z_3\big( ({\vec \Delta}_{1}\cdot {\vec \Delta}_{2})^2+ ({\vec \Delta}_{2}\cdot {\vec \Delta}_{3})^2 + ({\vec \Delta}_{3}\cdot {\vec \Delta}_{1})^2 \big) - 4 Z_4  \big(\vec{\Delta}_1 \cdot \vec{\Delta}_2 \times \vec{\Delta}_3\big)^2 + \cdots
\label{eq: free energy}
\end{eqnarray}
\end{widetext}
where $\alpha$
 is an inessential positive constant.

The quadratic term and the first quartic term in (\ref{eq: free energy}) set the overall magnitude of $\Delta^2 = \sum_i {\vec \Delta}_{i}^2$, but do not differentiate between different SDW states. The second quartic term in
 (\ref{eq: free energy}) determines whether SDW order develops only at one nesting vector, or at all three (depending on the sign of
 $Z_2-Z_1-Z_3$). Finally, the third quartic term and sixth order term  control the relative orientation of the vector order parameters, if SDW order develops at multiple wavevectors.
 Close to $T_N$ the expansion to order ${\vec \Delta}_{i}^4$ is generally sufficient, but we include the sixth order term because $Z_3$ is suppressed by an extra factor of $T_N/t_1$, which is exponentially small in the weak coupling limit. The relative smallness of $Z_3$  arises because
in the  integrals for $Z_1, Z_2$, and $Z_4$, all fermions can be simultaneously brought to the saddle points, whereas in the integral for $Z_3$, three fermions can be brought simultaneously to saddle points, but the remaining fermion stays far away from not only the saddle point but also the FS.

We evaluate the coefficients $Z_1$ -- $Z_4$
  to leading order in small $T_N/t_1$ and obtain~\cite{supplement}
 \begin{eqnarray}
 &&Z_1 = \frac{0.20 \log{\frac{t_1}{T_N}}}{\pi^4 T^2_c t_1}, ~~Z_2 = \frac{0.58}{\pi^4 T^2_N t_1}, \nonumber \\
&&Z_3 = -\frac{0.08}{\pi^2 T^2_N  t_1} \frac{T_N}{t_1},  ~~Z_4 = - \frac{0.1}{T^4_N t_1}
\label{dec_18_1}
\end{eqnarray}
The positivity of $Z_1$ guarantees a second order phase transition, with the type of SDW order depending on the signs and relative magnitudes of $Z_2$, $Z_3$,
 and $Z_4$. Since $Z_3$ is smaller by $T_N/t_1$ than $Z_{1,2}$, and
$Z_2$ is smaller by $\log \frac{t_1}{T_N}$ than $Z_1$, it follows that $Z_2-Z_1-Z_3<0$, so the system forms SDW order simultaneously at all three nesting
 vectors (the $3Q$ state).
 Meanwhile, the relative orientation of the three SDW order parameters
 is controlled by the sign of $Z_3$ at the smallest ${\bf \Delta}_i$, and by the sign of $Z_4$ at somewhat larger  ${\bf \Delta}_i$.
   Both $Z_3$ and $Z_4$ are negative and favor the non-chiral SDW order with the three ${\bf \Delta}_i$
    all aligned along the same axis.

    An order parameter of the form $\vec{\Delta} \bigl(e^{i\mathbf Q_{3}\cdot\mathbf r} + e^{i\mathbf Q_{1}\cdot\mathbf r} \pm e^{i\mathbf Q_{2}\cdot\mathbf r} \bigr)$ leads to spin moments on the lattice of the form shown in Fig.~\ref{fig: lattice}.
    A quarter of lattice sites have spin moment $3 \Delta$, the other three quarters have moment $-\Delta$.
  Such an order cannot be obtained from any spin Hamiltonian for local moments of constant magnitude on every site.
  Our result differs from earlier mean-field analysis~\cite{TaoLioriginal} which found non-coplanar insulating SDW order at weak coupling. We note, however,
   the  $3Q$
    state that we found, with non-equal spin length on different sites,  was not considered in that work and other earlier considerations of the type of SDW order.
  We found analogous results for fermions on a triangular lattice at Van Hove filling. This system is identical to graphene, except that
   the nesting is less robust and is spoiled already by
   second neighbor hopping.

\emph{Properties of a uniaxial SDW:} Is the uniaxial SDW state a metal or an insulator? To address this issue we need to compute the fermionic spectrum.
Without loss of generality, we take the SDW to be uniaxial along the $z$ axis, so that $S^z$ is a good quantum number, and spin-up and spin-down fermions decouple. Consider the state with ${\bf \Delta}_1 = {\bf \Delta}_2 ={\bf \Delta}_3 =\Delta\hat{\mathbf z}\,\sigma_3$. The up spins near the three Van Hove points are described by a simple $3\times3$ Hamiltonian
\begin{equation}
H = \left(\begin{array}{ccc} \epsilon_{1,\mathbf k} -\delta \mu & \Delta & \Delta \\ \Delta & \epsilon_{2,\mathbf k} - \delta \mu & \Delta \\ \Delta & \Delta & \epsilon_{3,\mathbf k} -\delta \mu \end{array} \right)
\label{m_1}
\end{equation}
where $\epsilon_1, \epsilon_2, \epsilon_3$ are the dispersions near the Van Hove points, Eq.~(\ref{n_1}), and
 $\delta \mu$
   is the SDW-induced shift of the chemical potential.
 The $3\times3$ Hamiltonian describing the spin down branch is obtained by taking $\Delta \rightarrow - \Delta$.
  At $\mathbf k =0$ (i.e., at Van Hove points) the energies of spin-up excitations
  $E_{\mathbf k} - \delta \mu$ are $-\Delta, -\Delta$, and $2\Delta$, and the energies of spin-down excitations are $\Delta, \Delta$, and $-2\Delta$.
 In conventional SDW states (e.g., SDW on a 2D square lattice)  $\delta \mu/\Delta \propto T_N/E_F$ is negligibly small and can be safely neglected.
 We find, however, that in our case $\delta \mu = -\Delta$, so that gapless excitations arise in the spin-down spectrum.

\begin{figure}
        \includegraphics[width = 0.99 \columnwidth]{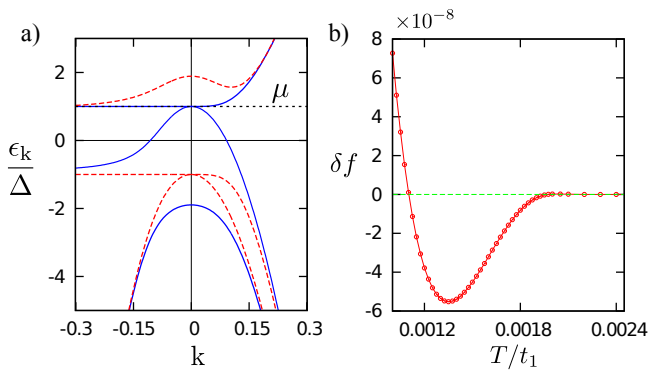}
     \caption{(Color online) a)
     Excitation spectrum
      $\epsilon_k = E_k-\delta \mu$
       of the $3Q$ uniaxial state. Negative $k$ are along the FS, positive $k$ are along
     the
     BZ boundary in the original BZ (along $k_x$ in the reduced zone).
           Placing the chemical potential at $\delta \mu = -\Delta$ ensures that
    four bands lie below the chemical potential (horizontal dotted line) and two lie above for all $\vec{k}$, irrespective of the value of $\Delta$. Thus the choice $\mu = - \Delta$ conserves electron number. Excitations with  spin projection
   opposite to  ${\bf \Delta}$ are in blue (solid), along ${\bf \Delta}$ are in red (dashed) lines. Note that
    gapless excitations arise in the spin-down branch only.
      b) Free energy difference $\delta F = F_{\rm uniaxial} - F_{\rm chiral}$ between the $3Q$ uniaxial SDW state and the chiral state, evaluated in the mean field approximation for the honeycomb lattice Hubbard model with $g_2 = g_3 = U = 1.7 t_1$ ($T_N \sim 0.002 t_1$).
       The $3Q$ uniaxial state has lower Free energy over a wide range of intermediate temperatures, but at the smallest $T$ the non-coplanar, chiral state, studied in earlier works~\cite{TaoLi,FaWang,Martin+Batista}, has lower Free energy.
        \label{fig: spectrum}}
     \end{figure}

 To see the unexpected shift of the chemical potential, we diagonalize Eq. (\ref{m_1}) and the corresponding equation for down spins and inspect six branches of excitations. We find that fixing $\delta \mu = -\Delta$ ensures that
   both in the paramagnetic and in the $3Q$ uniaxial SDW state
   there
 are
 four bands with $E_{\mathbf k} \leq \mu$ and two bands with $E_{\mathbf k} \geq \mu$ for all momenta in the reduced
  BZ
  (see Fig.~\ref{fig: spectrum}).
Since the chemical potential is fixed by the constraint that the total number of electrons (equal to the number of states below the chemical potential) must
 not change between $\Delta=0$ and $\Delta \neq 0$ \cite{altshuler}, it follows that we must set $\delta \mu = - \Delta$.  For verification, we computed the thermodynamic potential $\Omega (\Delta, \mu)$ from (\ref{eq: grand potential}), numerically solved the simultaneous equations
 $\partial \Omega/\partial \Delta =0$ and $\partial \Omega/\partial \mu = -N$, and confirmed that $\delta \mu = -\Delta$ to a high accuracy.

Having determined that $\delta \mu = -\Delta$, we find from (\ref{m_1}) that gapless excitations emerge when $\epsilon_{1,\mathbf k}\, \epsilon_{2,\mathbf k}\, \epsilon_{3,\mathbf k} =0$, which has solutions along three lines passing through each Van Hove point. Two of then coincide with the original FS, the third is directed towards the center of the
  BZ.
  The $3Q$ uniaxial SDW state is then obviously a metal. We  emphasize, however, that
 gapless states exist only for the electrons with spin projection
   opposite to
   ${\bf \Delta}$. The electrons with spin projection
    along  ${\bf \Delta}$  are fully gapped.  Since a Fermi surface exists for one spin projection only, we dub this state a `half metal.'
   We found an analogous `half-metal' spectrum
   for the $3Q$
    uniaxial
    SDW phase on the triangular lattice.

The half-metallic nature of the
SDW should manifest itself in numerous experiments. For example, in tunneling experiments conducted with electrons spin polarized along the $z$ axis, a hard gap will be seen for down spins, but a Fermi surface will be seen for up spins. Furthermore, since the low energy charged excitations involve up spins only, any charge currents will necessarily also be spin currents. Thus, the half metal state allows for electrical control of spin currents, which may be beneficial for nanoscience applications.

   \emph{Order parameter manifold:}~~~  The uniaxial SDW order obviously breaks $O(3)$ spin-rotational symmetry.
            It also breaks $Z_4$ discrete symmetry associated with either parallel or antiparallel ordering of ${\bf \Delta}_i$, i.e., in addition to the $(\Delta, \Delta, \Delta)$ state which
 we considered above, there are also $(\Delta, -\Delta, -\Delta)$,  $(-\Delta, \Delta, -\Delta)$, and  $(-\Delta, -\Delta, \Delta)$ states. These states have an identical structure of fermionic excitations, and correspond to the four in-equivalent ways to choose which two of the eight sites of the SDW unit cell carry large spins (see Fig. \ref{fig: lattice}(b)). Equivalently, the three other states from the $Z_4$ manifold are obtained from the $(\Delta, \Delta, \Delta)$ state by shifting the origin of coordinates to the center of one of three neighboring hexagons.
     An interesting possibility, which deserves further study, is that $Z_4$ symmetry can be broken before $O(3)$ symmetry, leading to a nematic-like state~\cite{fernandes}.

 \emph{The phase diagram:}~~~
Thus far we have constructed the Ginzburg-Landau expansion in small $\Delta/T_N$. This expansion
  becomes less justified
  as we move towards zero temperature. To investigate the behavior at
    arbitrary $T$
    we calculate
     numerically
     the full Free energies of the various SDW states from (\ref{eq: grand potential}).
     Upon doing this, we find that the $3Q$ uniaxial state
     has the lowest Free energy over a wide range of intermediate temperatures, but
          undergoes a first order transition at a lower temperature to the insulating chiral SDW state discussed in earlier works~\cite{TaoLi,FaWang,Martin+Batista}.
        We show the Free energy profile in Fig. \ref{fig: spectrum}b.
         We found this behavior both for graphene and for fermions on a triangular lattice.
Intuitively, the chiral SDW state wins at the lowest $T$  because it
 has spin-degenerate excitations and opens a full spectral gap, unlike the half-metal state.

 The Free energy profile in  Fig. \ref{fig: spectrum}b is for weak/moderate coupling, when $T_N/t_1 \ll 1$.
At $T_N\sim t_1$, the phase diagram is more complex and less universal.
 For completeness, we  discuss the forms of $Z_i$ and the phase diagram at $T_N \sim t_1$
  in the supplementary material \cite{supplement}.

{\it Conclusion:}~~~
 We considered in this work the SDW instability on the honeycomb and triangular lattices, when doped to the saddle points of the dispersion. The SDW instability is subleading to a $d-$wave
  superconducting instability at weak coupling, but becomes the leading instability if superconductivity is suppressed.
  We found that
  if the SDW ordering temperature $T_N$ is much smaller than the fermionic bandwidth,
then a uniaxial SDW order develops simultaneously at three inequivalent nesting vectors. This has an order parameter manifold $O(3) \times Z_4$ and corresponds to the ordering pattern shown in Fig.\ref{fig: lattice}. Such a state can only be obtained from an model of itinerant electrons with interactions, and not from a spin model of local moments. We found that such SDW state
is a half-metal in which gapless excitations exist in one spin branch only.
Such a state may be beneficial for nanoscience applications particularly because  charge currents will necessarily also be spin currents,  which allows for electrical control of the latter.

We thank L. Levitov for numerous discussions concerning the interplay between superconductivity and SDW order.
 We
 are also thankful to
 C. Batista, R. Fernandes, I. Martin, and Fa Wang
  for useful conversations.
G.W.C. is supported by ICAM and NSF-DMR-0844115, and A.V.C. is supported by  NSF-DMR-0906953.

\begin{widetext}
\section{Supplement}
In this supplement we present the calculations that were quoted in the main text.

\subsection{Calculation of $Z_1$}
We wish to evaluate
\begin{equation}
Z_1 = T \sum_{\omega_n} \int \frac{d^2k}{(2\pi)^2} G^2(\vec{k}, \omega_n) G^2(\vec{k+Q}_{3}, \omega_n)
\end{equation}
The integral over the Brillouin zone is dominated by those values of $\vec{k}$ where both Green functions correspond to states near a saddle point. Expanding the energy about the saddle points, we rewrite the integral as
\begin{equation}
Z_1 \approx T \sum_{\omega_n} \int \frac{d^2k}{(2\pi)^2} \frac{1}{\big(i\omega_n - \frac{3t_1}{4}(3k_x^2 - k_y^2)\big)^2\big(i\omega_n - \frac{3t_1}{4} 2k_y(k_y - \sqrt{3}k_x)\big)^2}
\end{equation}
Where the integral is understood to have a UV cutoff for $\vec{k}$ of order $1$.
We now define $a = \sqrt{3t_1/4} (k_y - \sqrt{3} k_x)$ and $b = \sqrt{3t_1/4}(k_y + \sqrt{3} k_x)$, and rewrite the above integral as
\begin{equation}
Z_1 = T \sum_{\omega_n} \frac{2}{3\sqrt{3} t_1}\int^{\sqrt{t_1}}_{-\sqrt{t_1}} \frac{da db}{(2\pi)^2} \frac{1}{\big(i\omega_n+ ab\big)^2\big(i\omega_n - a (a+b)\big)^2}
\end{equation}
We now define $x = ab$ and rewrite the integral as
\begin{equation}
Z_1 = T \sum_{\omega_n} \frac{2}{3\sqrt{3} t_1}\int^{\sqrt{t_1}}_{-\sqrt{t_1}} \frac{da}{2\pi} \frac{1}{|a|} \int_{-\sqrt{t_1} a}^{\sqrt{t_1}a} \frac{dx}{2\pi} \frac{1}{\big(i\omega_n + x\big)^2\big(i\omega_n - a^2 - x\big)^2}
\end{equation}
We now assume $T_N \ll t_1$ (which should certainly be the case for
 weak/moderate
 coupling). In this limit, we can perform the integral over $x$ approximately, using the Cauchy integral formula, to get
\begin{equation}
Z_1 = T \sum_{\omega_n} \frac{2}{3\sqrt{3} t_1}\int^{\sqrt{t_1}}_{-\sqrt{t_1}} \frac{da}{2\pi} \frac{1}{|a|} \frac{2 i \sign{\omega_n}}{(a^2 - 2 i \omega_n)^3} =  T \sum_{\omega_n} \frac{4}{3\sqrt{3} t_1}\int^{\sqrt{t_1}}_{-\sqrt{t_1}} \frac{da}{2\pi} \frac{1}{|a|} \frac{ i \sign{\omega_n (a^2 + 2 i \omega_n)^3}}{(a^4 + 4 \omega_n^2)^3}
\end{equation}
The imaginary part of the above integral is odd in $\omega$ and hence vanishes upon performing the Matsubara sum to leave an integral that is purely real
\begin{equation}
Z_1 = T \sum_{\omega_n} \frac{8 |\omega_n|}{3\sqrt{3} t_1}\int^{\sqrt{t_1}}_{-\sqrt{t_1}} \frac{da}{2\pi} \frac{1}{|a|} \frac{4 \omega_n^2 - 3 a^4}{(a^4 + 4 \omega_n^2)^3} \approx T \sum_{\omega_n} \frac{8 |\omega_n|}{3\sqrt{3} t_1}\int^{\sqrt{t_1}}_{-\sqrt{t_1}} \frac{da}{2\pi} \frac{1}{|a|} \frac{4 \omega_n^2}{(a^4 + 4 \omega_n^2)^3}
 \end{equation}
 with logarithmic accuracy. Performing the integral over $a$ (again with logarithmic accuracy) gives
 \begin{equation}
 Z_1 \approx T \sum_{\omega_n} \frac{1}{12 \pi \sqrt{3} t_1} \frac{1}{|\omega_n|^3} \ln \frac{t_1}{\omega_n} = \frac{1}{48 \pi^4 \sqrt{3} T_N^2 t_1} \big(16.8 \ln \frac{t_1}{2\pi T}  + 10.5 \big) \approx  \frac{16.8 \ln \frac{t_1}{T_N}}{48 \pi^4 \sqrt{3} T_N^2 t_1}% \big(16.8 \ln \frac{t_1}{2\pi T}  + 10.5 \big)
 \end{equation}
 Where we take $\omega_n = 2\pi (n+1/2) T_N$, $T = T_N$ and perform the discrete sum on mathematica.

 \subsection{Calculation of $Z_2$}
 This time we want to evaluate
 \begin{equation}
 Z_2 = T \sum_{\omega_n} \int \frac{d^2k}{(2\pi)^2} G^2(\vec{k}, \omega_n) G(\vec{k+Q}_{3}, \omega_n)G(\vec{k+Q}_{1}, \omega_n)
 \end{equation}
 Again, we anticipate this integral will be dominated by regions of the Brillouin zone where all three Green functions correspond to states near saddle points. Expanding the dispersion about the saddle points, we obtain
 \begin{equation}
Z_2 \approx T \sum_{\omega_n} \int \frac{d^2k}{(2\pi)^2} \frac{1}{\big(i\omega_n - \frac{3t_1}{4}(3k_x^2 - k_y^2)\big)^2\big(i\omega_n - \frac{3t_1}{4} 2k_y(k_y - \sqrt{3}k_x)\big)\big(i\omega_n - \frac{3t_1}{4} 2k_y(k_y + \sqrt{3}k_x)\big)}
\end{equation}
Making the same coordinate substitutions as in the preceding section, we recast this as
 \begin{eqnarray}
Z_2 = T \sum_{\omega_n} \frac{2}{3\sqrt{3} t_1}\int^{\sqrt{t_1}}_{-\sqrt{t_1}} \frac{da db}{(2\pi)^2} \frac{1}{\big(i\omega_n + ab\big)^2\big(i\omega_n - a (a+b)\big)\big(i\omega_n - b (a+b)\big)}%\\
\end{eqnarray}
After scaling out $\omega_m$, we can rewrite it as
 \begin{eqnarray}
Z_2 = T \sum_{\omega_n} \frac{2}{3\sqrt{3} t_1 |\omega_n| ^3}\int^{\sqrt{t_1}}_{-\sqrt{t_1}} \frac{da db}{(2\pi)^2} \frac{1}{\big(i+ ab\big)^2\big(i - a (a+b)\big)\big(i- b (a+b)\big)} = T \sum_{\omega_n} \frac{2}{12 \pi^2 \sqrt{3} t_1 |\omega_n| ^3} 2.9%\\
\end{eqnarray}
Where the rescaled integral is fully convergent, and can be done numerically on mathematica. The sum over Matsubara frequencies can also be done on mathematica, and yields the answer
 \begin{eqnarray}
Z_2  = T \sum_{\omega_n} \frac{2.9\times 16.8}{48 \pi^4 \sqrt{3} t_1 T_N^2}
\end{eqnarray}
Comparing with the previous expression for $Z_1$, we see that $Z_2 \approx Z_1 \times 2.9/\ln(T/t_1)$. Thus, $Z_2 \ll Z_1$ provided the log is large. (If the log is not large then the evaluation of $Z_1$ with logarithmic accuracy does not suffice, and sub-logarithmic contributions to $Z_1$ must also be taken into account.)

\subsection{Calculation of $Z_3$}
We want to evaluate
\begin{equation}
Z_3 = T \sum_{\omega_n} \int \frac{d^2k}{(2\pi)^2} G(\vec{k}, \omega_n) G(\vec{k+Q}_{3}, \omega_n)G(\vec{k+Q}_{1}, \omega_n) G(\vec{k+Q}_{1}+\vec{Q}_{3}, \omega_n)
\end{equation}
This time it is not possible to place all the Green functions at the saddle points. In fact, we cannot even place all the Green functions at the Fermi surface - the best that can be done is to place three of the Green functions near a saddle point, but the fourth has to be off Fermi surface. Thus, we obtain,
 \begin{equation}
Z_3 \approx T \sum_{\omega_n} \int \frac{d^2k}{(2\pi)^2} \frac{1}{\big(i\omega_n - \frac{3t_1}{4}(3k_x^2 - k_y^2)\big)\big(i\omega_n - \frac{3t_1}{4} 2k_y(k_y - \sqrt{3}k_x)\big)\big(i\omega_n - \frac{3t_1}{4} 2k_y(k_y + \sqrt{3}k_x)\big)(i \omega - 2 t_1)}
\end{equation}
Making the usual substitutions, and assuming $t_1 \gg T_N$, we obtain
 \begin{eqnarray}
Z_3 &\approx&  \sum_{\omega_n} \frac{T_N}{12 \sqrt{3} \pi^2 t_1^2 \omega_n^2} \int_{-t_1}^{t_1}  \frac{da db}{\big(i\omega_n+a b \big)\big(i\omega_n - a(a+b)\big)\big(i\omega_n -b(a+b)\big)} \\&\approx& \sum_{\omega_n} \frac{T_N}{12 \sqrt{3} \pi^2 t_1^2 \omega_n^2}   \int_{-t_1/\omega_n}^{t_1/\omega_n}  \frac{da db}{\big(i+a b \big)\big(i - a(a+b)\big)\big(i -b(a+b)\big)}
\end{eqnarray}
The integral is convergent. As usual, the imaginary part is odd in $\omega$ and vanishes and we care only about the real part. Performing the integral on mathematica and taking the real part, we obtain
\begin{eqnarray}
Z_3 &\approx& \sum_{\omega_n} \frac{6.5 T_N}{12 \sqrt{3} \pi^2 t_1^2 \omega_n^2}  \approx \frac{6.5}{48 \sqrt{3} \pi^2 t_1^2 T_N}
\end{eqnarray}
Which is parametrically smaller than $Z_1$ and $Z_2$ by $T_N/t_1$.

\subsection{Calculation of $Z_4$}
We now calculate the coefficient of the sixth order chirality sensitive term in the free energy, $v ({\bf \Delta}_{1} \cdot ({\bf\Delta}_{2} \times {\bf\Delta}_{3}))^2$. After some analysis of diagrams %(needs to be written up)
we find that,
\begin{equation}
Z_4 =  T \sum_{\omega_n} \int \frac{d^2k}{(2\pi)^2} G^2(\vec{k}, \omega_n) G^2(\vec{k+Q}_{3}, \omega_n)G^2(\vec{k+Q}_{1}, \omega_n)
\end{equation}
We can now place all the Green functions on the near a saddle point. Making the usual substitutions, we obtain
\begin{eqnarray}
 Z_4&=&  \sum_{\omega_n} \frac{T_N}{12 \sqrt{3} \pi^2 t_1 |\omega_n|^5} \int_{-t_1/\omega_n}^{t_1/\omega_n}  \frac{da db}{\big(i+a b \big)^2\big(i - a(a+b)\big)^2\big(i -b(a+b)^2\big)} \\
&=&   \sum_{\omega_n} \frac{T_N}{12 \sqrt{3} \pi^2 t_1 |\omega_n|^5} \int_{-t_1/T_N}^{t_1/T_N}  \frac{da db}{\big(i+a b \big)^2\big(i - a(a+b)\big)^2\big(i -b(a+b)^2\big)}
\end{eqnarray}
where in the second line we have assumed that the Matsubara sum is controlled by the first few Matsubara frequencies. The integral can be done numerically, and it is negative for small $T_N/t_1$. It then follows that  at weak/moderate coupling $Z_4<0$, so that the Free energy at sixth order also disfavors chirality.

\subsection{Scaling functions}
The above calculations were performed at weak/moderate coupling, assuming $T_N/t_1 \ll 1$. However, the integrals can be evaluated at arbitrary $T_N/t_1$. To this end, it is useful to define the scaling functions $f_i(T_N/t_1) = Z_i(T_N/t_1)/Z_i(0)$. These scaling functions are evaluated numerically and shown in Fig.\ref{fig: scaling functions}.

    \begin{figure}
a) \includegraphics[width = 0.3\columnwidth]{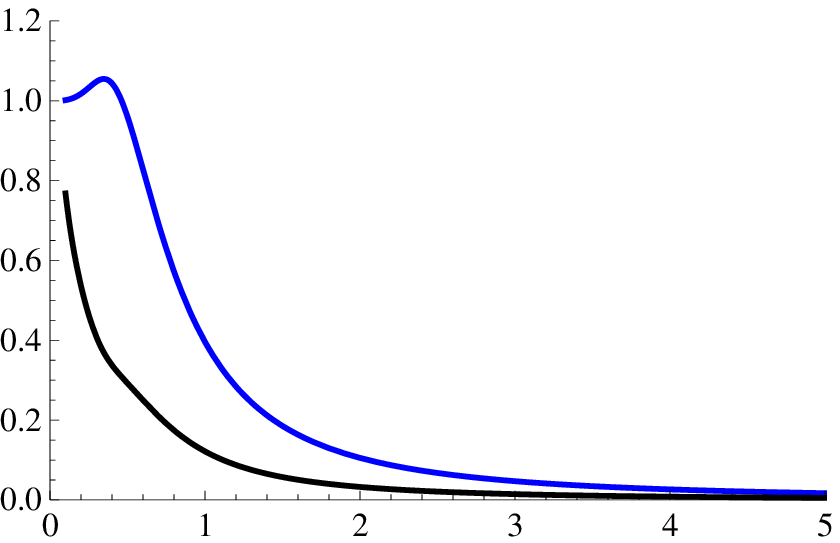}
b) \includegraphics[width = 0.3\columnwidth]{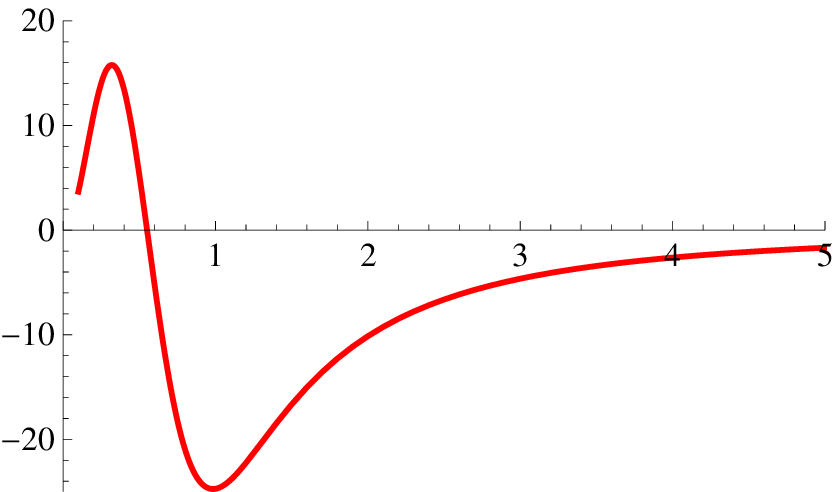}
c) \includegraphics[width = 0.3\columnwidth]{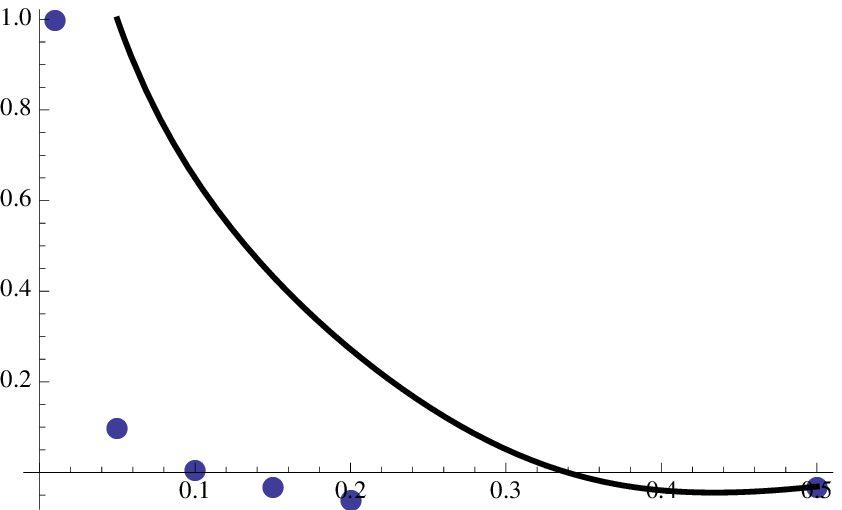}
\caption{(Color online) The behavior of scaling functions $f_i (x) = f_i \left(\frac{T_N}{t_1}\right)$. a)
 The scaling functions $f_1$ (black) and $f_2$ (blue). b) The
  scaling function $f_3$. c)
   The scaling function $f_5$ corresponding to the term $Z_2-Z_1-Z_3$ (solid line). Superimposed on this is a discrete plot of $f_4$ (points). Note that the scaling functions $f_3 (x)$, $f_4(x)$ and $f_5(x)$ change sign between small $T_N/t_1$ (weak/moderate coupling) and $T_N \sim t_1$ (strong coupling).\label{fig: scaling functions}}
\end{figure}

\section{Phase diagram at strong coupling}

At strong coupling, when $g_{2,3} \geq t_1$ and $T_N \sim t_1$,  our analysis based on Ginzburg-Landau expansion is less accurate because fermions can no longer be approximated as free particles (the self-energy corrections to fermionic lines and vertex corrections to square and hexagonal diagrams are generally of order one). Nevertheless, if we apply our analysis to $T_N \sim t_1$, we find that
 $Z_3$, $Z_4$ and $Z_2 -Z_1-Z_3$ all change signs at some $T_N/t_1$  (see Fig.~\ref{fig: scaling functions}).

The first sign change occurs in the sixth order chirality sensitive term $Z_4$, which becomes positive for $T_N/t > 0.1$. When $Z_4$ is positive, the chiral SDW state \cite{TaoLi} is energetically favored, provided we are sufficiently far below $T_N$ for the sixth order term to dominate over the quartic term $Z_3$. Thus, at large $T_N/t_1$, the uniaxial SDW phase has a much narrower region of stability, and the transition into the chiral SDW phase happens quite close to $T_N$.

The term $Z_1-Z_2-Z_3$ is next to change sign, becoming negative for $T_N/T> 0.35$. Once this term becomes negative, the system prefers instead a $1Q$ collinear state, of the form discussed in \cite{TaoLioriginal}, wherein SDW order develops only at a single nesting vector. The subsequent sign change of $Z_3$ at $T_N/t \approx 0.55$ has no physical consequences.

The $1Q$ collinear SDW state that forms at $T_N/t > 0.35$ is a (full) metal because the entire FS is not gapped out.  The competition between a metallic collinear state and non-coplanar insulating state has been detected numerically in the mean-field analysis at strong coupling \cite{Martin+Batista}, and our results for the strong coupling case are in line with this earlier study. Our strong coupling
     results are also consistent with the studies that found a non-coplanar, chiral SDW order in the models of spins of the same fixed length at
     every lattice site \cite{TaoLi, Martin+Batista}. However, in the weak/moderate coupling limit, our results indicate that the preferred state is a uniaxial $3Q$ state of a sort not considered before, which can only be realized in a model starting from itinerant fermions.

 We also analyzed the evolution of $Z_i$ with $T_N/t_1$ for  fermions on a triangular lattice. We found similar trends, e.g., sign change of $Z_4$. However, for a triangular lattice, the first sign change (in $Z_4$) occurs at a much larger $T_N/t_1 \sim0.5 $,  when the itinerant approach is very questionable.

\end{widetext}

\end{document}